\documentclass[a4paper,11pt]{iopart}
\usepackage{a4wide}
\usepackage[english]{babel}
\usepackage{graphicx}
\usepackage{iopams}
\usepackage{lscape}
\usepackage{xcolor}
\usepackage{accents}
\usepackage{tablefootnote}
\usepackage{geometry}
\usepackage{float}
\usepackage{dsfont}
\usepackage{braket}
\DeclareMathAlphabet{\mathpzc}{OT1}{pzc}{m}{it}

\eqnobysec

\begin{document}

\title{On the Majorana representation of the optical Dirac equation}
\author{Mark R. Dennis$^{1,2}$, Teuntje Tijssen$^{2}$ and Michael A. Morgan$^{3}$}
\address{$^1$ School of Physics and Astronomy, University of Birmingham, Edgbaston, Birmingham, B15 2TT, UK}
\address{$^2$ H.~H.~Wills Physics Laboratory, University of Bristol, Tyndall Avenue, Bristol, BS8 1TL, UK}
\address{$^3$ Physics Department, Seattle University, 901 12th Avenue, Seattle, WA 98122, USA}
\ead{mark.dennis@physics.org}

\begin{abstract}
We consider the representations of the optical Dirac equation, especially ones where the Hamiltonian is purely real-valued.
This is equivalent, for Maxwell's equations, to the Majorana representation of the massless Dirac (Weyl) equation.
We draw analogies between the Dirac, chiral and Majorana representations of the Dirac and optical Dirac equations, and derive two new optical Majorana representations.
Just as the Dirac and chiral representations are related to optical spin and helicity states, these Majorana representations of the optical Dirac equation are associated with the linear polarisation of light.
This provides a means to compare electron and electromagnetic wave equations in the context of classical field theory.
\end{abstract}


\section{Introduction}

It is now well understood that structured light beams have many analogies with structured wavepackets in quantum mechanics.
This has led to numerous advances to our understanding of light, such as accelerating Airy wavepackets, optical vortices, optical orbital angular momentum, and electromagnetic duality as a gauge transformation, as well as many other examples \cite{roadmap}. 
Michael Berry has contributed to many of these advances over the past five decades (e.g.~\cite{berry90,berry09}).

In particular, and perhaps as the best example of this analogy, the nature of optical spin has been explored by directly comparing the Maxwell equations with the Schr\"odinger-Pauli or Dirac equation for the spin-half electron. 
Thus a transverse, right or left circularly-polarised electromagnetic plane wave, as a spin $\pm 1$ state, corresponds to a stationary spin $\pm 1/2$ electron, or more literally, an ultrarelativistic electron with parallel spin and momentum directions (helicity states).

While this identification reveals significant properties of the spin angular momentum of light, it is guilty of a deficit in electromagnetic democracy: circular polarisation is no more fundamental than elliptical or linear polarisation.
What, therefore, of the analogy with the Dirac equation for linear optical polarisation?
We will address this question by comparing light beams with solutions of the Dirac equation in the \emph{Majorana representation} \cite{majorana}.
Such Majorana particle solutions, not appearing in the standard model of particle physics, are less familiar than electrons, although Majorana quasiparticles have recently received intense interest in condensed matter physics 
\cite{wilczek09,Beenakker13}.

As in many of Michael Berry's contributions to the study of the overlap between optics and quantum theory, we work in an entirely wave-mechanical (``first quantised'') setting.  
The source-free Maxwell equations $\partial_t\boldsymbol{E}=\nabla\times{\boldsymbol{B}}$, $\partial_t\boldsymbol{B}=-\nabla\times{\boldsymbol{E}}$, for divergence-free electric $\boldsymbol{E}$ and magnetic $\boldsymbol{B}$ fields, can be cast as a quantum-mechanical Schr\"odinger-Pauli/Dirac wave equation $\rmi \partial_t \psi = \widehat{H} \psi$ for an appropriate Hamiltonian operator $\widehat{H}$ and wavefunction $\psi$. 
We will directly compare solutions to these Maxwell equations with solutions to the Dirac equation for massless ``electrons'' (i.e., general spin-half fermions), which we claim can be identified as wavefunctions for a single quantum particle (``photon'', ``electron'' or ``Majorana particle''). 
We do not consider interactions between light and matter, and we use natural units such as Planck's constant $\hbar$ and the speed of light in vacuum $c$ have the value unity.

In the case of the Dirac equation, the wavefunction $\psi$ is a four-component Dirac spinor, and the Hamiltonian $\widehat{H} = \boldsymbol{\alpha}\cdot \widehat{\boldsymbol{p}} + m \alpha_0 = -\rmi \boldsymbol{\alpha}\cdot\nabla + m\alpha_0$ depends on a set of four $4\times 4$ Hermitian \emph{Dirac alpha matrices} $\alpha_0, \alpha_x, \alpha_y, \alpha_z$, $\boldsymbol{\alpha} = (\alpha_x, \alpha_y, \alpha_z)$ and the mass of the matter wave $m$. Here we've used the fact that the canonical momentum operator is $\widehat{\boldsymbol{p}} = -\rmi\nabla$. 
In this paper we deal exclusively with the case of massless electrons. Then, as we'll see in the next section, Maxwell's equations and the Dirac equation 
\begin{equation}
   \rmi \partial_t \psi = -\rmi \boldsymbol{\alpha}\cdot\nabla \psi,
   \label{eq:dirac1}
\end{equation}
(sometimes called the Weyl equation) share many structural features as systems of linear first-order partial differential equations in spacetime, with a multicomponent vector/spinor solution. 

Solutions $\psi$ for different representations of the Dirac matrices can be identified with different states of massive spin-half particles. 
In the so-called \emph{Dirac representation}, the spin operator $\widehat{S}_z$ is diagonal, so the basis states of $\psi$ are spin up and spin down, stationary electrons.
In the \emph{chiral representation} (also called the Weyl representation), high energy electrons traveling almost at light-speed (effectively massless), have basis states given by the helicity states with spin parallel or antiparallel to their propagation direction.

Finally, in the Majorana representation, the Dirac equation is completely real ($\boldsymbol{\alpha}$ are real and $\alpha_0$ is pure imaginary).
This is suited to describe Majorana solutions, which here we take to be eigenfunctions of the complex conjugation operator, which can be identified as purely real (eigenvalue $+1$) or purely imaginary (eigenvalue $-1$) \cite{hill:cpt,cmr97}.

Direct electromagnetic analogues of the Dirac equation (considered by Iwo Bialynicki-Birula \cite{ibb96}, Stephen Barnett \cite{barnett14}, see also \cite{horsley18,ktz18,sebens19}) emphasize that the Maxwell equations can be put into a form analogous to (\ref{eq:dirac1})
\begin{equation}
   \rmi \partial_t \mathcal{F} = -\rmi \bbeta\cdot\nabla \mathcal{F}
   \label{eq:maxwellanalog}
\end{equation}
where $\mathcal{F}$ is some 6-component \emph{Faraday spinor} whose entries are linear combinations of the components of electric and magnetic field vectors, and $\bbeta = (\beta_x,\beta_y,\beta_z)$ are an appropriate set of $6\times 6$ matrices, which in fact form a DKP (Duffin-Petiau-Kemmer) algebra \cite{umezawa,roman,duffin,kemmer36,kemmer39,kemmer43}, with matrices which, unlike a Clifford algebra, may be singular.

Bialynicki-Birula's ``photon wave function'' is the analogue of the chiral representation, involving $\mathcal{F}=(\bi{E}+\rmi \bi{B}, \bi{E}-\rmi \bi{B})^T$, demonstrating the analogy between helicity eigenstates of light and ultra-relativistic electrons/neutrinos.
On the other hand, Barnett's ``six-component spinor'' (also see Darwin \cite{darwin}) is the analogue of the standard Dirac representation, involving $\mathcal{F}=(\bi{E},\rmi \bi{B})^T$, and allows the comparison between an electron at rest and static $\bi{E}$ and $\bi{B}$ fields.

Of course, one difference between electrons and  light particles (``photons'') is that the latter are massless, and hence there is no need for $\beta_0$ (the analogue of $\alpha_0$) in (\ref{eq:maxwellanalog}).
Similarly, the massless Dirac equation (or ``{Weyl equation}'') requires only the three Pauli matrices $\balpha=(\sigma_x,\sigma_y,\sigma_z)$ as a sufficient set of Dirac matrices.

A Majorana particle (massless or not), being an eigenstate of complex conjugation, is always balanced between its positive and negative frequencies, just like the true representation of electromagnetic fields.

We will construct a representation of the three $\beta_i$ matrices analogous to Majorana's own construction for the $\alpha_i$ matrices to be purely imaginary.
We anticipate the corresponding spinor $\mathcal{F}$ will somehow pick out linear polarisations as privileged just as the other representations pick out circular polarisation.
For a better comparison, we should prefer a spherical basis rather than Cartesian, as this generalises to any spin (and is necessary for any representation of Dirac spinor).
We find that the choice of Majorana representation is not unique, with different choices not only revealing different aspects of the relation between linear and circular polarisation, but hopefully offering insight into the nature of Majorana solutions.

\section{On the analogy between the Maxwell and Dirac equations} \label{sec:rs}

We begin with a review of the standard Dirac wave equation for free massive electrons (spin-half fermions) \cite{umezawa,peskinschroeder,schwartz,weinberg1,dirac},
\begin{equation}
   \rmi \partial_t \psi =  -\rmi \boldsymbol{\alpha}\cdot\nabla \psi + m \alpha_0 \psi  =  \boldsymbol{\alpha}\cdot\widehat{\boldsymbol{p}}\,\psi + m \alpha_0 \psi = \widehat{H}\psi.
   \label{eq:dirac}
\end{equation}
For equation (\ref{eq:dirac}) to be appropriately relativistic (i.e., for it to remain form-invariant under Lorentz transformations), Dirac showed \cite{dirac} that the four Dirac matrices $\alpha_0, \alpha_x, \alpha_y, \alpha_z$ are required to satisfy the anticommutation relations of a Clifford algebra
\begin{equation} 
   \alpha_i^2 = {\tt1}, \qquad \alpha_i \alpha_j + \alpha_j \alpha_i = {\tt0}, \qquad i \neq j,\; i,j = 0,x,y,z,
   \label{eq:clifford}
\end{equation}
where ${\tt 1}$ is the identity matrix.
The fewest number of dimension for which four such matrices obey the algebra is four, so the Dirac matrices are $4\times 4$, and the Dirac spinors $\psi$ have four components. A particular realization of this algebra (\ref{eq:clifford}) in terms of complex-valued $4\times 4$ matrices is called a \emph{representation} of the Dirac algebra. As one often shows in a course on relativistic quantum mechanics, all representations of (\ref{eq:clifford}) are unitarily equivalent. 

The physical role of a representation of the Dirac matrices is related to which operators are diagonalised in that representation.
For example, in the \emph{Dirac representation} (where $\alpha_0$ is diagonal), the $z$-component of spin is diagonal, whereas in the \emph{chiral representation}, the helicity operator---the component of spin in the direction of momentum---is diagonal. In the \emph{Majorana representation}, the operator $(\partial_t + \rmi \widehat{H}^{\mathrm{M}})$ is required to be real-valued, which means the matrix elements of the matrices $\boldsymbol{\alpha}^{\mathrm{M}}$ are real, and the matrix elements of $\alpha_{\mathrm{0}}^\mathrm{M}$ are purely imaginary.
In this paper we refer to the Majorana representation of a wave equation as a `Majorana equation.' 
See Pal \cite{pal} for a review of Majorana and Weyl fermions.

For the special case of massless electrons with $m = 0$, (\ref{eq:dirac}) is sometimes referred to as a \emph{Weyl equation}, for the spinor $\varphi$
\begin{equation}
   \quad \rmi \partial_t \varphi = (\boldsymbol{\alpha}\cdot\widehat{\boldsymbol{p}}) \, \varphi.
   \label{eq:weyl}
\end{equation}
Now that the matrix $\alpha_0$ is not required, relativistic covariance of the Weyl equation only requires three independent, anticommuting $\alpha$ matrices. The standard set of Pauli matrices for $i=x,y,z$
\begin{equation} 
   \sigma_x = \left( \begin{matrix} 0 & 1 \\ 1 & 0 \end{matrix}\right), \qquad 
   \sigma_y = \left( \begin{matrix} 0 & \!\!\!-\rmi \\ \rmi & 0 \end{matrix}\right), \qquad
   \sigma_z = \left( \begin{matrix} 1 & 0 \\ 0 & \!\!\!-1 \end{matrix}\right),
   \label{eq:pauli}
\end{equation}
obey the Clifford algebra (\ref{eq:clifford}), so they suffice as a set of $\alpha_i$ matrices for the Weyl equation (\ref{eq:weyl}), and hence the \emph{Weyl spinors} $\varphi$ have only two components. 

The \emph{helicity operator} for a single quantum particle is defined as the projection of the spin along the direction of the particle's momentum or ${\boldsymbol{\Sigma}\cdot\widehat{\boldsymbol{p}}}/{|\boldsymbol{\Sigma}||\boldsymbol{p}|}$ where $\boldsymbol{\Sigma}$ is the spin operator, and $\widehat{\boldsymbol{p}}$ the momentum operator for the particle.
For a spin-half electron where $\boldsymbol{\Sigma}=\frac{1}{2}\boldsymbol{\sigma}$ the helicity operator is ${\boldsymbol{\sigma}\cdot\widehat{\boldsymbol{p}}}/{|\boldsymbol{p}|}$. The choice $\boldsymbol{\alpha}=\boldsymbol{\sigma}$ in (\ref{eq:weyl}) corresponds to Weyl spinors $\phi^+$ with positive helicity, whereas the choice $\boldsymbol{\alpha}=-\boldsymbol{\sigma}$, which still obeys (\ref{eq:clifford}), corresponds to  Weyl spinors $\phi^-$ with negative helicity. In other words, the massless spin-half Hamiltonian is $\pm|\boldsymbol{p}|$ times the helicity operator.

This shows the significance of the \emph{chiral representation} of the Dirac equation (\ref{eq:dirac}). In the massless case, the 4-component Dirac equation reduces to two uncoupled Weyl equations, one for each helicity.
Note that the Pauli matrices cannot be used as a set of $\alpha_i$ matrices for a massless Majorana equation, since $\sigma_y$ is purely imaginary (see \cite{aste} for exceptions.)

Next, we explain how to put the source-free Maxwell wave equations into a Hamiltonian form analogous to the Weyl (\ref{eq:weyl}) and massless Dirac (\ref{eq:dirac1}) wave equations. 
In an approach going back to Weber and Silberstein in the first decade of the twentieth century \cite{silb1,silb2,weber}, we begin by rewriting  the classical, source-free electromagnetic field as the complex \emph{Riemann-Silberstein vector} (RS vector)
\begin{equation}
   \boldsymbol{F}^{\pm} \equiv \boldsymbol{E} \pm \rmi \boldsymbol{B},
   \label{eq:RSv}
\end{equation}
for which Maxwell's equations become simply $\nabla \cdot \boldsymbol{F}^{\pm} = 0$ and
\begin{equation}
   \rmi \partial_t \boldsymbol{F}^{\pm} = \pm \nabla \times \boldsymbol{F}^{\pm},
   \label{eq:RSeq}
\end{equation}
called the \emph{Riemann-Silberstein (RS) equation}.
The RS vector therefore is divergenceless, and satisfies a Schr\"odinger-Pauli-like wave equation whose Hamiltonian $\widehat{H} = \pm \nabla \times$ is  first-order and real-valued. 
For this reason, $\boldsymbol{F}^{\pm}$ is sometimes identified as the `wavefunction' for electromagnetism \cite{ibb96}.
Furthermore, the continuous duality symmetry in the electromagnetic fields without sources, $\boldsymbol{E} \to \boldsymbol{E} \cos\tau + \boldsymbol{B}\sin\tau$, $\boldsymbol{B} \to -\boldsymbol{E} \sin\tau + \boldsymbol{B}\cos\tau$ for $0 \le \tau < 2\pi$, is a global gauge symmetry for the RS equation, since under this transformation $\boldsymbol{F}^{\pm} \to \boldsymbol{F}^{\pm} \rme^{\mp \rmi \tau}$, leaving the RS equation invariant \cite{ibb96,misnerwheeler,zwanziger,desteit,cambar,ifc13,bbn13}.

We can now express the RS equation (\ref{eq:RSeq}) in the form of a Schr\"odinger-Pauli wave equation by finding a Hamiltonian form for the curl operator. 
The cross product can be expressed by using a 3-vector of $3 \times 3$ spin matrices $\boldsymbol{S} = (S_x, S_y, S_z)$, where
\begin{equation}\fl\quad
   S_x = \left( \begin{matrix} 0 & 0 & 0 \\ 0 & 0 & -\rmi \\ 0 & \rmi & 0 \end{matrix}\right), \quad
   S_y = \left( \begin{matrix} 0 & 0 & \rmi \\ 0 & 0 & 0 \\ -\rmi & 0 & 0 \end{matrix}\right), \quad
   S_z = \left( \begin{matrix} 0 & -\rmi & 0 \\ \rmi & 0 & 0 \\ 0 & 0 & 0 \end{matrix}\right),
   \label{eq:Sdef}
\end{equation}
that is the components of $S_i$ are $(S_i)_{jk} = -\rmi \varepsilon_{ijk}$. 
So for arbitrary vectors $\boldsymbol{u}, \boldsymbol{v}$, we have $-\rmi (\boldsymbol{S}\cdot \boldsymbol{u})\,\boldsymbol{v} = \boldsymbol{u}\times\boldsymbol{v}$.
These matrices are the generators of rotations for vectors in a Cartesian basis around the $x,y,z$ axes respectively, satisfying the commutation relations $S_x S_y - S_y S_x = \rmi S_z$ and cyclic permutations. 

Since in terms of the momentum operator the gradient is $\nabla = \rmi\,\widehat{\boldsymbol{p}}$, we have that the RS Hamiltonian $\pm \nabla \times$ is $\pm$ the scalar product of the spin angular momentum with the momentum, $\pm \boldsymbol{S}\cdot \widehat{\boldsymbol{p}}$. For vector fields (i.e.~spin one electromagnetism), the spin operator is simply $\boldsymbol{\Sigma} = \boldsymbol{S}$.
Thus, the RS Hamiltonian is $\pm|\boldsymbol{p}|$ times the helicity operator, as was also true for the massless spin-half case.
Hence, the RS equation (\ref{eq:RSeq}) for $\boldsymbol{F}^\pm$ is equivalent to
\begin{equation}
   \rmi \partial_t \boldsymbol{F}^\pm = \pm (\boldsymbol{S}\cdot\widehat{\boldsymbol{p}}) \boldsymbol{F}^\pm.
   \label{eq:RS2}
\end{equation}
which is analogous to the Weyl equations (\ref{eq:weyl}) 
$\rmi\partial_t\varphi^\pm = \pm(\boldsymbol{\sigma}\cdot\widehat{\boldsymbol{p}})\varphi^\pm$ for massless electrons/neutrinos.

We've now shown that the 3-component RS vector wave-functions $\boldsymbol{F^\pm}$ for light are analogous to the 2-component Weyl spinors $\varphi^\pm$ for massless electrons. 
It's natural then to try to find a wave equation for light which is the analogue to the massless Dirac equation (\ref{eq:dirac1}). 
In fact, it turns out that in the early days of relativistic quantum mechanics, researchers followed similar reasoning, showing that the most general first order, Lorentz covariant wave equation $(\beta_\mu\partial^\mu+m)\psi=0$ for particles of mass $m$ and integral spin, requires \emph{beta matrices} (analogous to the Dirac gamma matrices) that obey the relation $\beta_\mu\beta_\nu\beta_\lambda+\beta_\lambda\beta_\nu\beta_\mu=\beta_\mu\delta_{\nu\lambda}+\beta_\lambda\delta_{\nu\mu}$ for all $\mu,\nu,\lambda=0,1,2,3$ \cite{umezawa,roman,kemmer36,duffin,kemmer39}. 
This defines the \emph{Duffin-Kemmer-Petiau algebra} (DKP algebra) of $\beta_\mu$ matrices. 
The DKP algebra for wave equations of integral spin replaces the Clifford algebra for the case of a wave equation with spin-half. 
Note that whereas matrices obeying the Clifford algebra can not be singular, there is no such requirement here. As for the Dirac equation, the massless case doesn't require the matrix $\beta_0$; we only need to work with the spatial components $\beta_i$. 
In this case the DKP algebra reduces to 
\begin{equation}
   \!\!\!\beta_i^3 = \beta_i, \quad\!\!
      \beta_i \beta_j \beta_i = {\tt0},  \quad\!\!
      \beta_i \beta_j^2 + \beta_j^2 \beta_i = \beta_i,  \quad\!\!
      \beta_i \beta_j \beta_k + \beta_k \beta_j \beta_i = {\tt0}  \quad\!\!
   \label{eq:DKP}
\end{equation}
for distinct spatial indices $i,j,k=1,2,3$. 
Note that the first and last equalities in (\ref{eq:DKP}) {\it are} satisfied by the Clifford algebra, but the middle two are not.
In general then, we seek a set of three such matrices, $(\beta_x, \beta_y, \beta_z) \equiv \boldsymbol{\beta}$ satisfying the \emph{optical Dirac equation}
\begin{equation}
   \rmi \partial_t \mathcal{F} \quad = \quad -\rmi(\boldsymbol{\beta}\cdot\nabla)\mathcal{F}\quad = \quad  (\boldsymbol{\beta}\cdot\widehat{\boldsymbol{p}})\mathcal{F},
   \label{eq:oDe}
\end{equation}
where the Faraday spinor $\mathcal{F}$ is a multicomponent wavefunction involving linear combinations of the components of the electric and magnetic field, which generalizes the RS vector wave-function (\ref{eq:RSv}). 
The optical Dirac equation (\ref{eq:oDe}) is the wave equation for light analogous to the massless Dirac equation for the electron.
Expression (\ref{eq:RS2}) satisfies this equation where $\boldsymbol{\beta} = \pm\boldsymbol{S}$ and $\mathcal{F} = \boldsymbol{F}^{\pm}$, exactly like the Weyl equation (\ref{eq:weyl}) satisfies the massless Dirac equation (\ref{eq:dirac1}) with $\boldsymbol{\alpha} = \pm\boldsymbol{\sigma}$ and $\psi=\varphi^\pm$.
In general, DKP algebras are analogous to the Dirac Clifford algebra for integer spin fields, and were historically introduced to describe spin zero and spin one meson fields, as closely as possible to the Dirac equation \cite{umezawa,roman,kemmer39,kemmer43,hc}.

Both the Dirac and the optical Dirac equations are covariant under proper Lorentz transformations. 
The rotation generators are given by the commutators of the $\alpha$ and $\beta$ matrices
\begin{equation}
   \Sigma^{\alpha}_i=-\frac{\rmi}{4}\varepsilon_{ijk}\alpha_j\alpha_k \quad\quad \Sigma^{\beta}_i=-\rmi\varepsilon_{ijk}\beta_j\beta_k, 
   \label{eq:Sigmadef}
\end{equation}
which are also the expressions for the spin angular momentum in the spin-half and spin-one cases respectively, and will be useful in what follows. 

\section{Representations of the Dirac and optical Dirac equations}\label{sec:reps}
In the representations considered here, the Dirac $\alpha$ matrices are $2+2$ block matrices, with the $\sigma_i$ matrices appearing in either diagonal or off-diagonal blocks.  
Similarly, according to the analogy between the optical Dirac equation and the Dirac equation discussed in the previous section, the  DKP $\beta$ matrices will be $3+3$ block matrices consisting of the $S_i$ matrices.
An efficient way of denoting such block matrices is to use the \emph{direct product} (or \emph{Kronecker product}) of two matrices, $C=A\times B$, where $C_{(a,j),(b,k)}\equiv A_{ab}\,B_{jk}$, so the rows and columns of the product matrix are labeled with pairs of indices corresponding to the original matrices. 
In our application here, matrix $A$ is taken from the set of Pauli matrices $\sigma_i$ where $i=0,1,2,3$, and $\sigma_0\equiv\tt 1$ is the $2\times2$ identity matrix. 
It turns out that the $4\times4$ Dirac $\alpha_j$ matrices may be expressed as a direct product $\sigma_i\times\sigma_j$, while the  $6\times6$ optical Dirac DKP $\beta_j$ matrices are expressed as $\sigma_i\times S_j$, the choice of $i=0,1,2,3$ depending generally on which representation of the $\alpha$ and $\beta$ matrices we choose, as we'll show explicitly below. 
Also, we'll use $\sigma_i\times\boldsymbol\sigma$ for the Cartesian vector of direct product matrices. 
The {\it mixed-product rule} $(A\times B)(C\times D)=(AC)\times(BD)$ follows directly from the definition of the direct product, and is particularly useful when computing products of $\alpha$ matrices or products of $\beta$ matrices, such as required in equations (\ref{eq:Sigmadef}) above, and (\ref{eq:gamma5}) below. 

First, we consider the massless Dirac equation, (\ref{eq:dirac1}). The simplest way of combining the two helicities into the Dirac equation is to work in the \emph{chiral representation},
\begin{equation}
   \boldsymbol{\alpha}^{\mathrm{c}} \equiv \sigma_3 \times\boldsymbol{\sigma} = \left( \begin{matrix} \boldsymbol{\sigma} & {\tt0} \\ {\tt0} & -\boldsymbol{\sigma} \end{matrix}  \right),
   \label{eq:alphaW}
\end{equation}
which is evidently the block diagonal (i.e.~direct sum) of the positive and negative helicity operators considered in the previous section. The Clifford algebra (\ref{eq:clifford}) is completed by $\alpha^{\mathrm{c}}_0 = \sigma_1\times\tt{1} = \left( \begin{smallmatrix} {\tt0} & {\tt1} \\ {\tt1} & {\tt0} \end{smallmatrix}\right)$. 
According to the first of equations (\ref{eq:Sigmadef}), the choice (\ref{eq:alphaW}) gives the spin operator as $\boldsymbol{\Sigma}^{\mathrm{c}} = {\tt1}\times\frac{1}{2}\boldsymbol{\sigma}=\frac{1}{2} \left( \begin{smallmatrix} \boldsymbol{\sigma} & {\tt0} \\ {\tt0} & \boldsymbol{\sigma} \end{smallmatrix}\right)$, which means that a rotation about the $z$-axis is diagonal in this representation. 

As discussed in the last section, the massless Dirac equation in chiral representation separates into two uncoupled equations for two Weyl 2-spinors $\varphi^\pm$, with  opposite helicity. (For this reason the chiral representation is often referred to as the {\it Weyl} representation.) The sign of the helicity is an eigenvalue of what here we call the \emph{chirality matrix} $X$,
\begin{equation}
   X \equiv -\rmi\,\alpha_x \alpha_y \alpha_z
   \label{eq:gamma5}
\end{equation}
which is easily seen to be the so-called fifth gamma matrix $\gamma^5 = \rmi \gamma^0\gamma^1\gamma^2\gamma^3$ ($\gamma^0\equiv\alpha_0,\gamma^i\equiv\alpha_0 \alpha_i$) \cite{peskinschroeder,schwartz,weinberg1}, valid in any representation.
In the chiral representation, $X^{\mathrm{c}} = \sigma_3\times \tt{1} =  \left( \begin{smallmatrix} \tt{1} & \tt0 \\ \tt0 & -\tt{1} \end{smallmatrix}\right)$, so the 4-spinor solution to the Dirac equation $\psi^{\mathrm{c}}$ is composed of Weyl spinors of positive and negative helicities, $\psi^{\mathrm{c}} = ( \varphi^+ ,\, \varphi^-)^T$. We note here that many QFT text books have the reverse ordering (upper components with negative helicity), corresponding to the opposite signs in (\ref{eq:alphaW}) and $X^{\mathrm{c}}$ \cite{peskinschroeder,schwartz}.

The matrices representing chirality $X^{\mathrm{c}}$ and $z$-component of spin $\Sigma_z^{\mathrm{c}}$ commute, with the four components of $\psi^{\mathrm{c}}$ corresponding, in turn, to positive helicity, positive spin; positive helicity, negative spin; negative helicity, positive spin; negative helicity, negative spin.
For a positive-energy plane wave corresponding to a massless particle with momentum $p=E>0$ in the $+z$ direction, only positive helicity positive spin corresponding to the spinor $\psi^{\mathrm c}\sim\left(1,\,0,\,0,\,0\right)^T\rme^{\rmi(p z - E t)}$, and negative helicity negative spin with spinor $\psi^{\mathrm c}\sim\left(0,\,0,\,0,\,1\right)^T\rme^{\rmi(p z - E t)}$, are possible.
Mixing signs of spin and helicity is only possible for negative energy plane waves. 

In the \emph{Dirac representation} the $\alpha$ matrices are defined in terms of the direct product with $\sigma_1$, namely
\begin{equation}
   \boldsymbol{\alpha}^{\mathrm{D}} \equiv \sigma_1\times\boldsymbol{\sigma} = \left( \begin{matrix} \tt0 & \boldsymbol{\sigma} \\ \boldsymbol{\sigma} & \tt0 \end{matrix}  \right),
   \label{eq:alphaD}
\end{equation}
with $\alpha_0^{\mathrm{D}} = \sigma_3\times \tt{1} = \left( \begin{smallmatrix} \tt{1} & \tt0 \\ \tt0 & -\tt{1} \end{smallmatrix}  \right)$, and (\ref{eq:gamma5}) gives the chirality operator $X^{\mathrm{D}} = \sigma_1\times \tt{1}= \left( \begin{smallmatrix} \tt{0} & \tt1 \\ \tt1 & \tt{0} \end{smallmatrix}  \right)$. As in the chiral representation, (\ref{eq:Sigmadef}) gives $\boldsymbol{\Sigma}^{\mathrm{D}} = \tt{1}\times\frac{1}{2}\boldsymbol{\sigma}$, so again $\Sigma_z$ is diagonal.
This is the original representation used by Dirac and is frequently used to describe massive electrons, especially in the non-relativistic limit. Such a positive-energy electron plane wave is an eigenstate of spin only if it has zero momentum, given by the spin up and spin down Dirac spinors $(1,\,0,\,0,\,0)^T$ and $(0,\,1,\,0,\,0)^T$ (the other two basis spinors correspond to zero momentum negative energy states for the electron). 
The two positive energy massless plane waves discussed in the last paragraph, with positive and negative helicities, correspond in this representation to the Dirac spinors $\psi^{\mathrm{D}}\sim(1,\,0,\,1,\,0)^T \rme^{\rmi(p z - E t)}$ and  $(0,\,1,\,0,\,-1)^T \rme^{\rmi(p z - E t)}$.

We defer discussing the \emph{Majorana representation} of the Dirac equation until the next section, where we work in a spherical basis.

We turn now to representations for the optical Dirac equation (\ref{eq:oDe}). By direct analogy with the Dirac equation the {\it chiral representation} for the optical Dirac equation is
\begin{equation}
   \boldsymbol{\beta}^{\mathrm{c}} \equiv \sigma_3\times\boldsymbol{S} = \left( \begin{matrix} \boldsymbol{S} & \tt0 \\ \tt0 & -\boldsymbol{S} \end{matrix}\right),
   \label{eq:betaW}
\end{equation}
which corresponds to the positive helicity operator in the upper left block, and the negative helicity operator in the lower right block. The reader can easily check that these $\beta$ matrices obey the DKP algebra (\ref{eq:DKP}). 
The 6-dimensional Faraday spinor in the chiral representation is given in terms of the RS vectors (\ref{eq:RSv}) by
$\mathcal{F}^{\mathrm{c}} = (\boldsymbol{F}^+ ,\, \boldsymbol{F}^-)^T$ (appropriately normalized). 
The upper and lower components of the optical Dirac equation in chiral representation are the $\pm$ RS equations in (\ref{eq:RS2}). 
This is the form of the optical Dirac equation constructed in Bialynicki-Birula \cite{ibb96} (and also by Majorana \cite{majorana2}, also see \cite{oppenheimer31,archibald55,good57,fradkin66,mrb74}). 
The chirality operator is clearly $X^{\mathrm{c}} = \sigma_3\times \tt{1} =  \left( \begin{smallmatrix} \tt{1} & \tt0 \\ \tt0 & -\tt{1} \end{smallmatrix}\right)$ (where $\tt1$ is the $3\times3$ identity matrix) as with the chiral representation for the Dirac equation, although there is no $\beta$ matrix analogue to (\ref{eq:gamma5}). 
According to equation (\ref{eq:Sigmadef}), the spin operator is $\boldsymbol{\Sigma}^{\mathrm c}=\tt1\times\boldsymbol{S}$. 
Notice that since the 3-vectors here are in the Cartesian basis $\Sigma_z^{\mathrm c}= \left( \begin{smallmatrix} S_z & \tt0 \\ \tt0 & S_z \end{smallmatrix}\right)$ is not diagonal. 
We consider the spin basis in the next section. 

This chiral representation is the natural one for describing circularly polarized light. 
The Faraday spinors corresponding to circularly polarized plane waves with momentum $p=E>0$ traveling in the $+z$ direction can be easily found by solving the RS equations (\ref{eq:RS2}). 
The solution for the case of positive helicity (R-handed light) is $\mathcal{F}^{\mathrm{c}} \sim (\boldsymbol{e}_+, \boldsymbol{0})^T \,\rme^{\rmi(p z - E t)}$, and for the case of negative helicity (L-handed light) is $\mathcal{F}^{\mathrm{c}} \sim (\boldsymbol{0}, \boldsymbol{e}_-)^T \,\rme^{\rmi(p z - E t)}$, where $\boldsymbol{e}_h$ are the eigenvectors of the helicity operator $S_z$ with eigenvalues $h=\pm1$. 
Note the correspondence to the plane wave solutions for the massless Dirac equation in chiral representation given above. 
There, instead of the 3-vectors $\boldsymbol{e}_h$, we have spin-up $\left(\begin{smallmatrix} 1\\0\end{smallmatrix}\right)$ and spin down $\left(\begin{smallmatrix} 0\\1\end{smallmatrix}\right)$ Weyl 2-spinors for the case of positive and negative helicity, respectively. 
A feature of this chiral description is that the duality gauge symmetry is simply a phase transformation generated by $-\rmi X^{\mathrm{c}}$. 

Moving on to the {\it Dirac representation} (or Darwin representation) for the optical Dirac equation and  following the same logic as above, we have
\begin{equation}
   \boldsymbol{\beta}^{\mathrm{D}} \equiv \sigma_1\times\boldsymbol{S} = \left( \begin{matrix} 0 & \boldsymbol{S} \\ \boldsymbol{S} & 0 \end{matrix}  \right),
   \label{eq:betaD}
\end{equation}
with Faraday spinor $\mathcal{F}^{\mathrm{D}} = (\boldsymbol{E} ,\,\rmi \boldsymbol{B})^T$, which was discussed at length by Barnett \cite{barnett14} (also see \cite{mohr10}). 
Unlike the chiral representation of the optical Dirac equation, the electric and magnetic field now occur in separate components of the spinor.
We recover the RS equations (\ref{eq:RS2}) in this representation by adding and subtracting the upper and lower components of the Faraday tensor.

It is interesting to note that a static, irrotational electric field $\boldsymbol{E}$ satisfies Maxwell's equations with $\boldsymbol{B} = 0$, which is a counterpart to the case of a stationary electron being in a spin eigenstate.
For the case of circularly polarized positive energy plane waves discussed previously, the Faraday spinors are found to be $\mathcal{F}^{\mathrm{D}} \sim (\boldsymbol{e}_+, \boldsymbol{e}_+)^T \,\rme^{\rmi(p z - E t)}$ and $\mathcal{F}^{\mathrm{D}} \sim (\boldsymbol{e}_-, -\boldsymbol{e}_-)^T \,\rme^{\rmi(p z - E t)}$ respectively, for the case of positive helicity and negative helicity. This again is completely analogous to the electron result in the Dirac representation, as given above. 

Note that in this representation, $X^{\mathrm{D}} = \sigma_1 \times {\tt 1}$, and $-\rmi X^{\mathrm{D}}\mathcal{F}^{\mathrm{D}} = (\boldsymbol{B},\,-\rmi\boldsymbol{E})^T$ also satisfies the optical Dirac equation.
Just as in the chiral representation, $-\rmi X^{\mathrm{D}}$ generates the electromagnetic duality transformation.
Additionally, one can show that $\mathcal{F}^{\mathrm{c}}$ and $\mathcal{F}^{\mathrm{D}}$ behave the same way under spatial rotations, and in both cases $\boldsymbol{\beta}\cdot\delta\boldsymbol{v}$ generates boosts by a velocity $\delta\boldsymbol{v}$.

Finally, since we have considered $\beta$ matrices constructed as direct products $\sigma_3\times\boldsymbol{S}$ and $\sigma_1\times\boldsymbol{S}$, it is natural to consider the final case of $\sigma_2\times\boldsymbol{S}$.
In this case, the $\beta$ matrices become purely real, so (\ref{eq:oDe}) becomes a purely real equation, which can therefore be considered a {\it Majorana representation} in the Cartesian basis.
It is convenient to include a $-$ sign in the definition of the $\beta$ matrices, 
\begin{equation}
   \boldsymbol{\beta}^{\mathrm{M}} \equiv -\sigma_2\times\boldsymbol{S} = \left( \begin{matrix} 0 &  \rmi \boldsymbol{S} \\ -\rmi \boldsymbol{S} & 0 \end{matrix}  \right),
   \label{eq:betaM}
\end{equation}
so that the Faraday spinor is $\mathcal{F}^{\mathrm{M}} = (\boldsymbol{E} ,\, \boldsymbol{B})^T$.
This representation, which is  very similar to the Dirac representation above, has been considered previously in the literature \cite{darwin,inagaki94,sz}, but not explicitly as a Majorana equation. 
The Faraday spinor $\mathcal{F}^{\mathrm{M}}$ is purely real when $\boldsymbol{E}$ and $\boldsymbol{B}$ are, giving classical electromagnetism with only real numbers; underpinning the fact that electromagnetism (massless spin one boson) is trivially Majorana.
Finally, we note that duality transformations here are realized by the direct product of the 2D rotation $\left(\begin{smallmatrix} \cos\tau&\sin\tau\\\!\!\!-\sin\tau&\cos\tau \end{smallmatrix}\right)$ matrix with the $3\times3$ identity matrix, generated by $-\rmi X^{\mathrm{M}} = \rmi \sigma_2 \times {\tt 1}$. 

\section{Spin basis representation and the optical Majorana equation}\label{sec:sb}

We have so far considered vectors in the Cartesian basis: 3-vectors were expressed in a basis of real, mutually orthogonal vectors $\boldsymbol{e}_x$, $\boldsymbol{e}_y$, $\boldsymbol{e}_z$.
The representation $\boldsymbol{\beta}^{\mathrm{M}}$ is not equivalent to Majorana's representation of the Dirac equation, however, since $\boldsymbol{\beta}^{\mathrm{M}}$ gives real matrices in the Cartesian representation, whereas Majorana was working in the spin basis.
In this section, we consider the complex spin basis (spherical basis) $\boldsymbol{e}_+$, $\boldsymbol{e}_0$, $\boldsymbol{e}_-$, of eigenvectors of $S_z$ (with eigenvalues $+1$, $0$, $-1$ respectively.
The unitary matrix transforming from the Cartesian basis to the spin basis $U = \tfrac{1}{\sqrt{2}}\left( \begin{smallmatrix} 1 & -\rmi & 0 \\ 0 & 0 & \sqrt{2} \\ 1 & \rmi & 0 \end{smallmatrix} \right)$, such that
\begin{equation}
   U \left( \boldsymbol{e}_x, \boldsymbol{e}_y, \boldsymbol{e}_z \right)^T = \left( \boldsymbol{e}_+, \boldsymbol{e}_0, \boldsymbol{e}_- \right)^T,
   \label{eq:Ucs}
\end{equation}
and the inverse $U^{-1} = U^{\dagger}$.
It should be noted this matrix slightly differs from its counterpart in quantum spin physics, as the standard optical convention does not quite follow the Condon-Shortley convention (which would define right-handed circular polarisation as $-\boldsymbol{e}_+$).

We can now write down the $S$ matrices in the spin basis, written $S^{\mathrm{s}}$, by conjugating with respect to these unitary matrices, $\boldsymbol{S}^{\mathrm{s}} = U\boldsymbol{S}U^{\dagger}$,
\begin{equation}\fl
   S^{\mathrm{s}}_x = \frac{1}{\sqrt{2}}\left( \begin{matrix} 0 & -1 & 0 \\ -1 & 0 & 1 \\ 0 & 1 & 0 \end{matrix}\right), \quad
   S^{\mathrm{s}}_y = \frac{1}{\sqrt{2}}\left( \begin{matrix} 0 & \rmi & 0 \\ -\rmi & 0 & -\rmi \\ 0 & \rmi & 0 \end{matrix}\right), \quad
   S^{\mathrm{s}}_z = \left( \begin{matrix} 1 & 0 & 0 \\ 0 & 0 & 0 \\ 0 & 0 & -1 \end{matrix}\right).
   \label{eq:Ssdef}
\end{equation}
With this choice, of course, $S^{\mathrm{s}}_z$ becomes diagonal (and real), with $S^{\mathrm{s}}_x$ real and $S^{\mathrm{s}}_y$ pure imaginary, just like the Pauli matrices.
These real and imaginary assignments are always necessary since in the angular momentum algebra formalism for any spin, raising and lowering operators are given by $\tfrac{1}{\sqrt{2}}(S^{\mathrm{s}}_x \pm \rmi S^{\mathrm{s}}_y)$, which must be real-valued matrices.  

Considering the Dirac equation, one can follow Majorana's construction \cite{majorana} of a real representation of the $\alpha$ matrices by choosing real $\alpha_x$ and $\alpha_z$ analogously to an existing representation, such as the Dirac representation, $\alpha^{\mathrm{M}}_x = \sigma_1\times\sigma_x$, $\alpha^{\mathrm{M}}_z = \sigma_1\times\sigma_z$.
The Clifford algebra rules (\ref{eq:clifford}), together with the restriction to real values, give exactly one choice for $\alpha_y^{\mathrm{M}}$, namely $\alpha_y^{\mathrm{M}} = \sigma_3\times\tt{1} = \alpha_0^{\mathrm{D}}$.
Furthermore, $\alpha_0^{\mathrm{M}}$ must be purely imaginary in the Majorana representation, and is therefore equal to the Dirac representation $\alpha_y^{\mathrm{D}}$.
Equation (\ref{eq:gamma5}) gives the chirality operator as $X^{\mathrm{M}}= \sigma_3\times\sigma_y$. 
Clearly, this Majorana representation is not unique; it would have been just as possible to start with the analogous chiral representation for $\alpha_x^{\mathrm{M}}=\sigma_3\times\sigma_x$ and $\alpha_z^{\mathrm{M}}=\sigma_3\times\sigma_z$, for which then the unique, real-valued  $\alpha_y^{\mathrm{M}}$ is $\alpha_y^{\mathrm{M}} = \sigma_1\times\tt{1} = \alpha_0^{\mathrm{c}}$, and it follows that $\alpha_0^\mathrm{M}=\alpha^\mathrm{c}_y$ and $X^{\mathrm{M}}=\sigma_1\times\sigma_y$.

Following the analogous procedure for the $\beta$ matrices, we find the DKP algebra has a purely real representation 
\begin{equation}
   \beta^{\mathrm{sM1}}_x = \sigma_1\times S^{\mathrm{s}}_x, \qquad \beta^{\mathrm{sM1}}_z = \sigma_1\times S^{\mathrm{s}}_z, \qquad \beta^{\mathrm{sM1}}_y = \sigma_2\times S^{\mathrm{s}}_y,
   \label{eq:betasM1}
\end{equation}
where here we've assumed the form of the first two matrices, and then used the DKP algebra ({\ref{eq:DKP}}) to find a unique, real-valued $\beta^{\mathrm{sM1}}_y$.
The most general special unitary transformations which gives these $\beta$ matrices (choosing to transform from the Cartesian basis Majorana representation $\boldsymbol{\beta}^{\mathrm{M}}$ defined in Section \ref{sec:reps}) is a one-parameter group, parameter $\tau$ results in a Faraday spinor with components 
\begin{eqnarray}
\fl   \mathcal{F}^{\mathrm{sM}1} & = & \tfrac{1}{\sqrt{2}} \left\{ \left( E_x + E_y ,\, \sqrt{2} E_z ,\, E_x - E_y ,\, -B_x + B_y ,\, -\sqrt{2} B_z ,\, -B_x - B_y \right)^T \!\!\cos\tau \right.\nonumber \\
\fl & & + \left. \left(B_x + B_y ,\, \sqrt{2} B_z ,\, B_x - B_y  ,\, E_x - E_y ,\, \sqrt{2} E_z  ,\,E_x + E_y \right)^T \!\!\sin\tau \right\}.
   \label{eq:FsM1}
\end{eqnarray}
Here the freedom of the similarity transformation reproduces the duality gauge symmetry.
Alternatively, with the choice $\tau=0$, we get $-\rmi X^{\mathrm{M}1}$ acting on the first spinor in (\ref{eq:FsM1}) gives the second spinor.

We now make some observations.
Although we started from the spin basis, the first and third components here are those of the Cartesian components of $\boldsymbol{E}$ in the basis $\boldsymbol{e}_{\nearrow} = (\boldsymbol{e}_x + \boldsymbol{e}_y)/\sqrt{2}$, 
$\boldsymbol{e}_0 = \boldsymbol{e}_z$ and $\boldsymbol{e}_{\searrow} = (\boldsymbol{e}_x - \boldsymbol{e}_y)/\sqrt{2}$, i.e.~the standard Cartesian basis rotated $45^{\circ}$ around the $z$-axis, and similarly for the fourth and sixth components involving the magnetic field. Strikingly, despite ostensibly working in the spin basis, this is a real, Cartesian basis, and $\boldsymbol{e}_{\nearrow}, \boldsymbol{e}_{\searrow}$ form a third mutually unbiased basis, together with $\boldsymbol{e}_{x}, \boldsymbol{e}_{y}$ and  $\boldsymbol{e}_{+}, \boldsymbol{e}_{-}$ in the transverse plane. Therefore, as expected, enforcing a Majorana-like condition on the optical Dirac equation in the spin basis has given us linear polarisation, as anticipated, but in a different basis from the original Cartesian basis $\boldsymbol{e}_x, \boldsymbol{e}_y$.

Alternatively, we can choose $\beta_x$ and $\beta_z$ from the chiral representation to construct a second Majorana-like real representation in the spin basis,
\begin{equation}
   \beta^{\mathrm{sM2}}_x = \sigma_3\times S^{\mathrm{s}}_x, \qquad \beta^{\mathrm{sM2}}_z = \sigma_3\times S^{\mathrm{s}}_z, \qquad \beta^{\mathrm{sM2}}_y = \sigma_2\times S^{\mathrm{s}}_y,
   \label{eq:betasM2}
\end{equation}
and now the most general Faraday spinor is
\begin{eqnarray}
\fl   \mathcal{F}^{\mathrm{sM}2} & = &  \tfrac{1}{\sqrt{2}} \left\{ \left( E_x + B_y ,\, \sqrt{2} E_z ,\, E_x - B_y ,\, -E_y - B_x,\, -\sqrt{2} B_z ,\, E_y - B_x \right)^T \!\!\cos\tau \right. \nonumber \\
\fl & & \left. \!+ \left(-E_y + B_x ,\, \sqrt{2} B_z ,\, E_y + B_x ,\, E_x - B_y ,\, \sqrt{2} E_z ,\, E_x + B_y \right)^T \!\!\sin\tau \right\}.
   \label{eq:FsM2}
\end{eqnarray}
Based on $\sigma_3$, this representation involves combinations of electric and magnetic field components similar to helicity states, but in a new way with only real coefficients, sharing the real-valuedness of $\boldsymbol{E}$ and $\boldsymbol{B}$.
Setting $\tau = 0$ initially, we can see that the first entry combines $E_x$ and $B_y$ -- which are equal for a linearly polarised electromagnetic wave traveling in $+z$, as are $E_y$ and $-B_x$ from the last entry.
On the other hand, $E_x - B_y$ and $E_y + B_x$ both must vanish for a linearly polarised plane wave in $+z$, regardless whether the field is real-valued or complex-valued.
The first and last components vanish for a field propagating in $-z$, again irrespective of whether fields are real- or complex-valued.
Therefore, instead of picking \emph{helicity} eigenstates from $z$-propagating plane waves, this second spin basis Majorana representation detects the \emph{direction of linear momentum}, based on the orientation of the electric and magnetic fields!

Elliptic or circular polarisation in the $xy$-plane has the same zero components in entries 2,3,4,5 of the Faraday spinor, but both the first and last entries are now complex numbers; similar to helicity, an arbitrary polarisation of a plane wave with a given direction is a superposition of orthogonal linear polarisations (as here), or opposite circular polarisations.
The duality angle $\tau$ here rotates the basis choice of these electric-magnetic components around the $z$-axis, a symmetry always broken on choosing a direction of linear polarisation.

\section{Discussion}\label{sec:dis}

We have explored various representations of the optical Dirac equation involving the integral spin DKP algebra, and found a strong analogy to the Dirac equation for the massless electron.
In fact, since electromagnetic fields (and, upon second quantization, photons) are intrinsically real, they are automatically their own antiparticles in the formalism.
Maxwell's equations therefore display a greater similarity to the Majorana representation of the massless Dirac equation, than to the Dirac equation for a massive electron.

The main feature was two new Majorana representations, one with Faraday spinor components being linear polarisation components, but in the mutually unbiased linearly polarised basis (in the transverse plane); the other roughly analogous to the helicity representation, but again favouring real superpositions of electric and magnetic fields which might vanish for certain linear polarisations.

In the standard optics framework, monochromatic fields lead to time dependence being suppressed, and fields are often represented by their complex-valued, positive frequency parts.
These naturally resemble complex-valued quantum matter fields, which are complex in the standard quantum formalism due to the positive mass and nonzero charge of the electron, although photons are massless and charge-free.
Thus, in the quantum formalism, classical electromagnetic fields are real-valued, for which it is easier to represent linear polarisation.

We have avoided giving a physical interpretation to negative-frequency waves. 
Of course, any observable effects properly manifest through the energy density of the field, which is provided by the $00$ component of the fields' relativistic energy-momentum tensor. 
The expression of this tensor is fundamentally different depending on whether the field is integer or half-integer \cite{pauli}, although there are analogues in the separation of spin and orbital parts \cite{berry09}.

The main historical reason for the development of the DKP algebra used here was to describe massive spin one fields (i.e.~satisfying the Proca equation), and a natural extension of this work would be to build an understanding of the effect of including mass in electromagnetic-like phenomena.
However, including an extra $\beta_0$ matrix to the DKP algebra significantly changes the representation, and the Dirac equation-like analogue to the Proca equation, whose field quantity is the analogue of the 4-potential rather than $\boldsymbol{E}$ and $\boldsymbol{B}$, requires $10\times 10$ DKP matrices, significantly larger than anything discussed here.
It would be nevertheless interesting to explore Weyl, Dirac, and Majorana-like representations of this more complicated DKP algebra (and the limit, in this algebra, $m \to 0$ \cite{hc}).

Much of the current interest in Majorana particles is focused towards quasiparticle excitations in condensed matter physics (e.g.~\cite{kitaev01,fukane08,LeFl12,Alicea12,Nadj-Perge14}, especially associated with topologically protected edge state modes.
Photonic analogues of various of these have been proposed (e.g.~\cite{bsn15,barim,rodmoya,xu}).
It would potentially be interesting to take the case of the optical Majorana equation into materials and interfaces, especially as the natural $s$ and $p$ polarisations associated with reflection and refraction at the edges of dielectric materials are linear polarisations, i.e.~Majorana-like states.
The true exotic nature of Majorana particles---their Majorana mass---may yet arise in an interpretation of the behavior of light in a suitable medium.

For free-space electromagnetism, however, our analysis merely emphasizes the fact that we can choose the basis in which we describe electromagnetic phenomena.
The circular or helical bases are analogous to the chiral and Dirac representations of electrons, and so emphasize spin and angular momentum aspects.
We've shown here that the basis states of linear polarization are analogous to the Majorana representation, which we hope may strengthen the power of analogies between different areas of physics.

\ack
This research was supported by the EPSRC and the Leverhulme Trust Research Programme RP2013-K-009, SPOCK: Scientific Properties of Complex Knots.

\end{document}